# Photoluminescence Blinking beyond Quantum-Confinement: Spatiotemporally Correlated Intermittency over Entire Micron Sized Perovskite Polycrystalline Disks

Nithin Pathoor[1,#], Ansuman Halder[2,#], Amitrajit Mukherjee[1], Jaladhar Mahato[1], Shaibal K. Sarkar[2,]*, and Arindam Chowdhury[1,3,]*

[1]Department of Chemistry, and [2]Department of Energy Science and Engineering, Indian Institute of Technology Bombay, Powai, Mumbai 400076, India

*Email: AC: arindam@chem.iitb.ac.in; SKS: shaibal.sarkar@iitb.ac.in

(Dated: 22 August 2018)

**Abstract:** Abrupt fluorescence intermittency or blinking is long recognized to be characteristic of single nano-emitters. Extended quantum-confined nanostructures also undergo spatially heterogeneous blinking, however, there is no such precedence in dimensionally unconfined (bulk) materials. Here, we report multi-level blinking of entire individual organo-lead bromide perovskite micro-crystals (volume 0.1-3 μm³) under ambient conditions. Extremely high spatiotemporal correlation (>0.9) in intra-crystal emission intensity fluctuations signifies effective communication amongst photogenerated carriers at distal locations (up to ~4 μm) within each crystal. Fused polycrystalline grains also exhibit this intriguing phenomenon, which is rationalized by correlated and efficient migration of carriers to a few transient non-radiative traps, the nature and population of which determine blinking propensity. Observation of spatiotemporally correlated emission intermittency in bulk semiconductor crystals opens up the possibility to design novel devices involving long range (mesoscopic) electronic communication.

Fluorescence intermittency or blinking, which refers to temporally random discrete jumps in intensity between bright and dark levels, has been considered as one of the foremost evidence for the detection of single nano-sized quantum-emitters.[1–8] Apart from single molecules, blinking is commonly observed in various individual quantum-confined systems such as semiconductor nanocrystals (NCs) where excitons/charge carriers are spatially restricted in more than one dimension (D).[4–13] In NCs, photoluminescence (PL) blinking is attributed to intermittent Auger ionization-recombination processes leading to charging-discharging of NCs or long lived carrier trapping in surface (defect) states.[4,5,14,15] However, PL intermittency is seldom observed beyond the nanoscale (approaching bulk), as temporally uncorrelated intensity fluctuations from various emitters average out over the ensemble, and contribution of surface states in radiative recombination becomes less significant compared to that of free carriers in the bulk.[7]

Even for 1- or 2-D confined extended nanostructures, blinking beyond diffraction limit (~250 nm) is uncommon, and such PL intermittency is spatially heterogeneous (*i.e.*, spatiotemporally uncorrelated).[9,16] There is a rare example of spatially concerted PL intensity fluctuations in an extended quantum-confined system;[10] a small proportion (1-2%) of entire single CdSe quantum-wires were found to exhibit correlated multi-level blinking, attributed to delocalized 1-D excitons which allow efficient long-range carrier migration. More recently, individual stacked-monolayers of transition metal dichalcogenides ($MoSe_2$/$WS_2$) were reported to undergo temporally *anti*-correlated blinking arising from mobile 2-D excitons which experience sporadic inter-layer charge transport.[17]

Owing to their exceptional carrier diffusion lengths and diffusivities,[18,19] organo-metal (hybrid) perovskites have gained considerable attention as suitable materials for photovoltaics. Hybrid halide perovskite (HHP) nanorods (250 nm in length) have been shown to undergo spatially extended multi-level blinking induced by metastable defects which efficiently trap photogenerated carriers in close vicinity.[11,12] Recently, single-crystal nanorods of HHPs were found to exhibit spatially extended heterogeneous blinking, attributed to mobile charged traps which transiently quench the emission of the surroundings even beyond the diffraction limit.[20] A similar mechanism is likely to be responsible for temporally uncorrelated PL blinking in nanodomains of doped HHP polycrystalline grains.[21]

In a quest to explore whether larger HHP crystals can exhibit spatially extended PL intermittency, we investigated methylammonium (MA) lead halide microcrystals (MCs) where efficient long range carrier diffusion has been reported.[18,19] Here, we present an extraordinary phenomenon where individual micron size MA lead bromide crystals undergo spatiotemporally-correlated PL blinking as single entities. For bulk materials without any dimensional confinement, the observation of spatially-synchronous blinking is unprecedented, and suggests long range communication between photogenerated carriers spatially separated far beyond the diffraction-limit.

$CH_3NH_3PbBr_3$ ($MAPbBr_3$) MCs were grown on glass coverslips using a solution processed method, where density of crystals could be controlled by precursor concentration and spin casting rate. The details of synthetic and experimental procedures are provided in Supporting Information (SI). X-Ray Diffraction (XRD) and Transmission Electron Microscopy (TEM) data (Figure S1, SI) corroborate perovskite structure with polycrystalline grains. Scanning Electron Microscopy (SEM) image of a typical sample area (Figure 1a) shows formation of quasi-circular disks, however, a few grains are sometimes fused together to form elongated MCs. Atomic Force Microscopy (AFM) measurements (Figure S2, SI) reveal that the grains resemble a mesa hundreds of nanometers in height, indicating a dimensionally unconfined bulk material. The solid-state optical spectra (Figure S3, SI) show an excitonic absorption feature at ~520 nm (2.38 eV) and an intense emission peak at ~544 nm (2.28 eV), while the average fluorescence lifetime ($\langle\tau\rangle$) is ~51 ns (Figure S4, SI). To investigate spatiotemporal PL behaviours of MCs, we performed epifluorescence video microscopy (at 25 Hz) under ambient conditions (295K), where samples were excited using a 405 nm laser, and the collected emission was imaged using a CCD camera (details provided in SI).

Fluorescence imaging of $MAPbBr_3$ MCs (Figure 1b) reveal spatially non-uniform emission from individual grains, with higher intensity at edges compared to that at interiors, similar to recent reports on other perovskites.[21–23] Such contrasting behaviour likely owes to difference in electronic structure and radiative recombination dynamics at the boundaries and interior regions of MCs. More interestingly, we find that PL intensity of each MC fluctuates dramatically with time (see Movie M1, SI), exemplified in Figure 1c using time-lapse image snapshots for two MCs (MC-1 and MC-2). The PL intensity trajectories for entire $MAPbBr_3$ MCs (Figure 1d) reveal sudden jumps between various dim and bright levels on top of a slow (<1Hz) time varying base intensity. The recurrence of high frequency (>10Hz) prominent PL fluctuations is validated using intensity jump-distributions for each MC (Figure 1e), which have significantly larger width compared to that for the background noise. We designate abrupt (40-80 ms) intensity-jumps with amplitudes more than five times the standard deviation of the background fluctuations (~4 cts) as blinking (Figure S5, SI). In stark contrast to $MAPbBr_3$, all-inorganic perovskite (CsPbBr$_3$) MCs of comparable dimensions do not blink at all (Figure 1d and Figure S6, SI). It is emphasized that no two segregated MAPbBr$_3$ MCs in the same movie undergo correlated PL intermittency (Figure 1d and Figure S7, SI), ruling out external factors such as laser intensity variations as a cause. Neither does blinking arise from temporally random spectral shifts (Figure S8, SI), nor is it a consequence of pronounced intensity fluctuations only within certain nanodomains, suggestive of spatial homogeneity in blinking dynamics across each crystal.

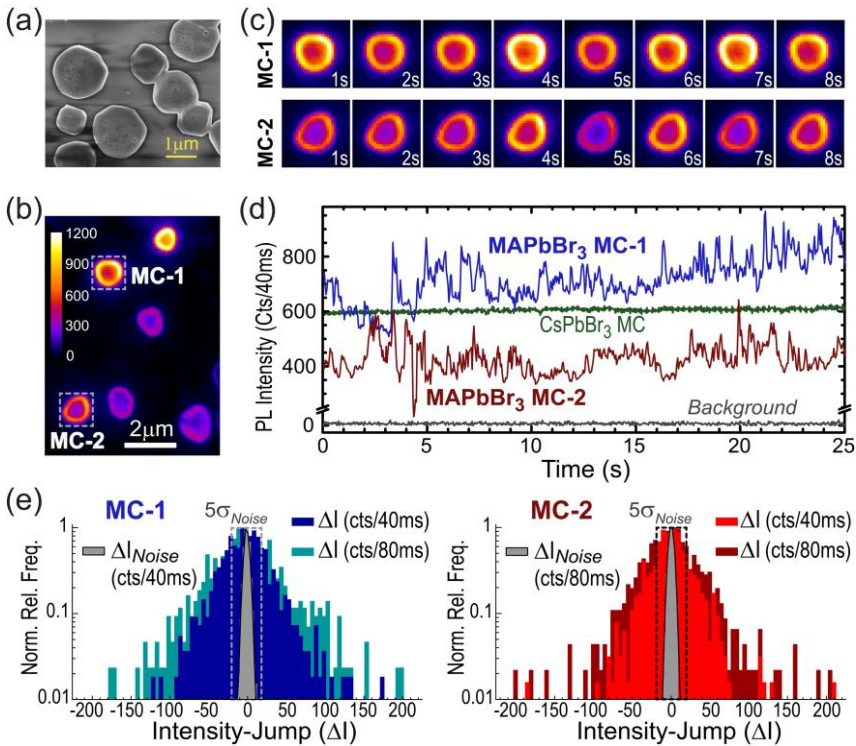

**Figure 1.** PL intermittency of entire isolated MCs. Typical (a) SEM and (b) PL image of $MAPbBr_3$ MCs on a glass coverslip. (c) Sequential time-lapsed (1s) PL image snapshots of two individual crystals, MC-1 and MC-2 (see Movie M1, SI), (d) Integrated PL intensity trajectories of entire MC-1 and MC-2 along with emission intensity fluctuations for an entire single $CsPbBr_3$ MC and that for the background noise. (e) Distribution of emission intensity-jumps (per 40 and 80 ms) for MC-1 and MC-2 along with that for the background (see Figure S5, SI).

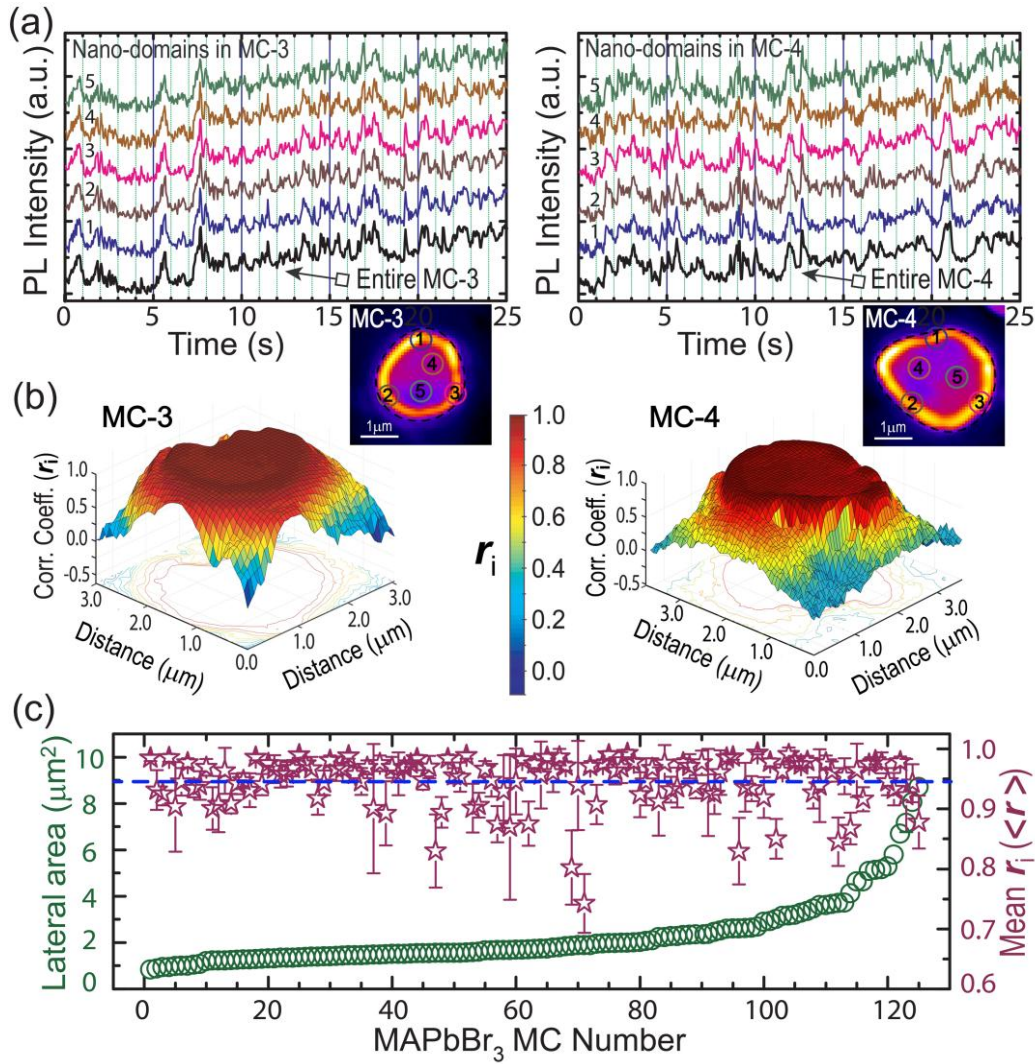

**Figure 2.** Spatially-correlated blinking in individual MCs. (a) Normalized PL intensity fluctuations from five nano-domains of two MAPbBr$_3$ MCs (*3* and *4*, insets) and the entire MC traces (see Movies M2-M3, SI). (b) Spatial maps of correlation coefficients ($r_i$) for MC-3 and MC-4. (c) Spatially-averaged intra-crystal $r$, $\langle r \rangle$ (stars) and corresponding size (circles) of 125 isolated MCs. Vertical bars depict standard deviation of $r$ for each MC. Horizontal line represents mean value of the distribution (see Figure S10, SI).

A comparison of spatially-resolved PL trajectories within two larger MCs is shown in Figure 2a. These reveal nanodomains within each MC exhibit nearly indistinguishable dynamical trend, although blinking amplitudes vary considerably over different spatial locations (Figure S9, SI). Importantly, temporal overlay of peaks and troughs for the spatially-resolved trajectories as well as the respective entire MC, provides compelling evidence that individual crystals undergo correlated blinking as a whole. To quantify spatially-synchronous PL fluctuations of each MC, we evaluated the Pearson correlation coefficient, $r$, (see methods, SI) between intensity trajectories at every location ($i^{th}$ pixel) within an MC and that for the entire entity. Using $r$ values for each pixel ($r_i$), we generated a spatial map of correlation coefficients for individual MCs (Figure 2b). These correlation images show extremely high $r_i$ values (typically >0.85) over entire MCs (Figure S9, SI), and sharply fall off to ~0.2 beyond boundaries. We have analyzed correlation images of 125 single MAPbBr$_3$ MCs of lateral dimensions between 0.8-9 μm² (Figure 2c), and found that an overwhelming majority (>95%) have a spatially averaged $r$ ($\langle r \rangle$) over 0.9 with nominal standard deviations. From the statistical distribution constructed using 125 isolated MCs (Figure S10, SI), the ensemble average of $\langle r \rangle$ was found to be 0.95±0.04, which further ascertains incredibly high intra-crystal spatiotemporal correlation in PL fluctuations for nearly all MCs up to lateral dimensions of ~10 μm².

Spatially synchronous PL blinking of entire micron sized (bulk) crystals is an unambiguous indicator of extremely long-range (> μm) communication amongst carriers photogenerated at diverse spatial locations, which is astounding. This implies that despite MAPbBr$_3$ being a semiconductor, a large number of carriers are extensively delocalized within each crystal, which in effect, determines their collective migration over large distances. Our inference is consistent with high charge carrier diffusivities (~$10^8$ μm²/s) and diffusion lengths (a few micrometers) reported for HHPs.[18,24] Further, rarely do we find spatially-correlated blinking beyond 5 μm in larger crystals, indicative of a strong connection between entire MC intermittency and carrier diffusion propensity.

In this context, it is relevant to note that grain boundaries can often hinder carrier diffusion owing to potential barriers created by structural defects.[25] To probe the effect of grain boundaries, we inspected relatively large, elongated MAPbBr$_3$ MCs with conjoined grains (see Figure 1a). The PL image of such an MC is shown in Figure 3a, where four fused grains (*I-IV*) can be readily identified via their prominent edges. Spatially-integrated PL trajectories (Figure 3b, *top*) reveal that all these grains collectively blink in unison (Movie M4, SI). However, segregated grains (such as *V*) in close vicinity does not (Figure 3b, bottom). Concerted blinking of fused grains (*I-IV*) is validated by very high inter-grain correlation coefficient (0.85-0.93), in contrast to nominal value (0.22) for disjointed grains (Figure 3c). Moreover, spatial correlation in blinking progressively withers, albeit slightly, with inter-grain distance, indicating marginal decrease in communication amongst carriers in grains located farther from one another.

To evaluate the spatial variation in the extent of correlated blinking, we plotted the temporal evolution of PL intensity along a strip spanning the fused MC (Figure 3d). We find that abrupt intensity fluctuations always occurs in-sync over the entire strip although slow temporal modulations in base intensity are not necessarily uniform. Using the spatially-resolved intensity trajectories, we generated a Pearson correlation matrix ($r_{ij}$) for each pair of pixels (*i* and *j*) (see methods, SI), where off-diagonal elements represent extent of inter-pixel blinking correlation (Figure 3e). High $r_{ij}$ values (>0.8) corroborate that majority of nanodomains in the strip blink in-concert. Intriguingly, the blinking-correlation does not diminish uniformly with distance between pixels; there are few localized zones for which correlation is modest ($r_{ij}$ ~0.6±0.1). This is primarily due to non-uniform slow (> sec) intensity modulations at certain local regions rather than the occurrence of spatiotemporally incoherent blinking events. We find that majority (~60%) of fused MC grains undergo concerted blinking over extended regions (up to ~10 μm$^2$), while remaining ones exhibit sporadic spatially-uncorrelated blinking (Figure S11, SI), the precise reasons for which still remain elusive.

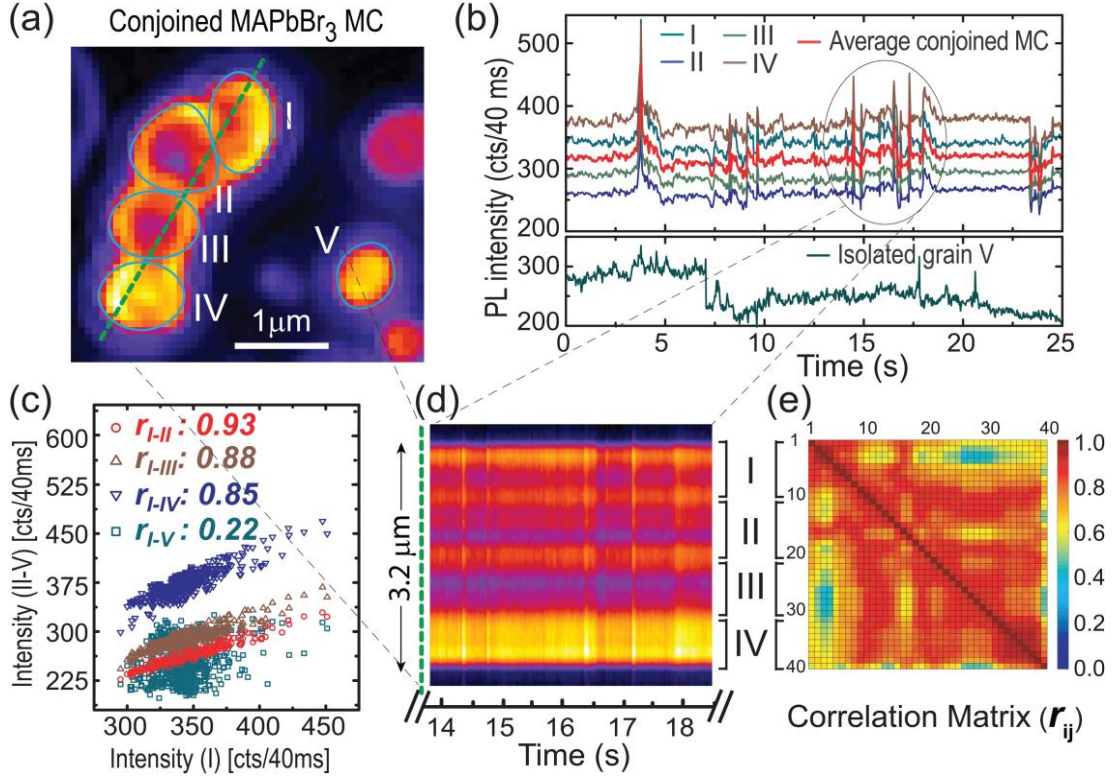

**Figure 3.** Concerted blinking of conjoined MC grains. (a) An elongated MAPbBr$_3$ MC with fused grains (*I-IV*) and a disjointed grain (*V*) nearby. (b) Synchronous blinking dynamics of *I-IV* and entire fused MC (top) in contrast to *V* (bottom) (see Movie M4, SI). (c) Integrated intensities of *II-V* against *I* for each frame and corresponding correlation coefficients. (d) Spatially-resolved intensity trajectory and (e) correlation coefficient matrix for each pair of pixels, along the 3.2 μm long strip traversing the fused MC grains.

While there are variations in blinking dynamics amongst individual MCs in the ensemble, the frequency of PL intermittency is nominally affected by excitation powers (0.1-10 Wcm$^{-2}$) or photon energy (405-532 nm). It is therefore unlikely that blinking in bulk crystals involves Auger ionization or surface trapping of charges. Here, we propose a phenomenological model for multi-level blinking of entire MAPbBr$_3$ MCs. First, prominent PL intensity fluctuations involve creation and removal of defects which can temporarily quench the emission in their vicinity.[2,11] The absence of spectral diffusion upon blinking (Figure S6, SI) suggests that metastable defects are non-radiative (NR) energetic funnels (traps). Second, spatially correlated intensity fluctuations in MAPbBr$_3$ must intricately relate to effective communication amongst majority of carriers throughout entire MCs. Extensive delocalization of charge-carrier wavefunctions[10] and/or strong correlation amongst photogenerated carriers is likely to play a significant role in such communication processes. In effect, carriers generated at any location can recognize the transiently formed NR traps within an MC, and high carrier diffusivities (~10$^8$ μm$^2$/s) in MAPbBr$_3$ permit fast (~100s of ns) carrier migration to distal (~μms) NR traps. As a consequence, the PL emission for entire MC grains is abruptly quenched. Upon annihilation or passivation of metastable traps, carriers generated at various locations are unable to access certain NR channels, and thus, PL intensity is (momentarily) augmented throughout individual MCs. Since prominent PL intensity fluctuations of entire crystals occur at ~100 ms, the process of NR trap formation and annihilation ought to take place at comparable timescales, much slower than that for correlated migration of carriers.

To verify whether entire MC blinking involves mesoscale carrier migration, we probed the PL emission from conjoined MCs via regioselective excitation. We find significant emission emanates from entire unilluminated (adjacent) fused-grain (Figure S12, SI), which provides concrete evidence that carriers can migrate over few micrometers before recombining radiatively. It is important to note that, due to shorter average radiative-recombination lifetimes ($\langle \tau \rangle$ ~51 ns) (Figure S4, SI) compared to the timescales of correlated carrier migration, entire MC PL blinking always occurs on top of a dominant base intensity level. Moreover, multi-level intermittency with a wide distribution of fluctuation amplitudes is likely to arise from a few metastable traps, each of which can partially quench the emission of an entire single MC to different extents.[11] We surmise that spatiotemporal variation in NR trap population leads to diverse blinking dynamics for different MCs in the ensemble, as well as for each MC at different time-windows. We have preliminary evidence that the metastable NR traps, at least in part, originate from adsorption/desorption of environmental constituents such as moisture; measurements performed under controlled environments reveal that entire MC blinking is initiated only above a threshold relative humidity (~30%) and is augmented in the presence of oxygen.[26] While further studies are necessary to completely elucidate entire MC blinking - indentify the exact nature of metastable quenchers as well as the specific role of organic (MA) cations, our results hopefully, will stimulate others to propose alternate plausible mechanisms.

To conclude, we report a unique phenomenon of spatially correlated fluorescence blinking of entire individual MCs of MAPbBr$_3$. Our results demonstrate that polycrystalline bulk materials can also undergo multi-level blinking, which dispels the long-standing notion that nanoscale carrier-confinement is essential to exhibit fluorescence intermittency. Spatial synchronicity of intra-crystal blinking has profound implications; this is a clear indicator of extremely long-range (~ few microns!) communication amongst majority of photogenerated carriers, which can potentially be harnessed for novel device or sensory applications.

We propose that entire crystal blinking originates from mesoscopic correlated migration of charge carriers to a few metastable non-radiative traps, however, a comprehensive understanding of the mechanism is necessary. Visualization of intra-microcrystal concerted intermittency opens up an avenue to estimate lower bounds of carrier diffusion parameters in bulk perovskites as well as other dimensionally unconfined semiconductor microstructures.

**Acknowledgements:** Financial support provided by Solar Energy Research Institutes of India and US (SERIIUS) and Ministry of New and Renewable Energy (MNRE), Govt. of India. A.C. thanks the National Center for Photovoltaic Research and Education (NCPRE), aided by MNRE for partial financial support. Authors acknowledge IRCC and CRNTS, IIT Bombay for usage of central facility equipment. N.P. and A.M. acknowledge fellowships from CSIR (India). A.C. thanks Subhabrata Dhar, Siuli Mukhopadhyay, Amitabha Tripathi, and the Kaleidoscope-2017 meeting for valuable discussions.

[1] X. S. Xie, R. C. Dunn, *Science* **1994**, *265*, 361–364.
[2] D. A. Vanden Bout, W. Yip, D. Hu, D. Fu, T. M. Swager, P. F. Barbara, *Science* **1997**, *277*, 1074–1077.
[3] R. M. Dickson, A. B. Cubitt, R. Y. Tsien, W. E. Moerner, *Nature* **1997**, *388*, 355–358.
[4] M. Nirmal, B. O. Dabbousi, M. G. Bawendi, J. J. Macklin, J. K. Trautman, T. D. Harris, L. E. Brus, *Nature* **1996**, *383*, 802–804.
[5] M. Kuno, D. P. Fromm, H. F. Hamann, A. Gallagher, D. J. Nesbitt, *J. Chem. Phys.* **2001**, *115*, 1028–1040.
[6] M. Lippitz, F. Kulzer, M. Orrit, *ChemPhysChem* **2005**, *6*, 770–789.
[7] J. Cui, A. P. Beyler, T. S. Bischof, M. W. B. Wilson, M. G. Bawendi, *Chem. Soc. Rev.* **2014**, *43*, 1287–1310.
[8] A. Swarnkar, R. Chulliyil, V. K. Ravi, M. Irfanullah, A. Chowdhury, A. Nag, *Angew. Chem. Int. Ed.* **2015**, *54*, 15424–15428.
[9] V. V. Protasenko, K. L. Hull, M. Kuno, *Adv. Mater.* **2005**, *17*, 2942–2949.
[10] J. J. Glennon, R. Tang, W. E. Buhro, R. A. Loomis, *Nano Lett.* **2007**, *7*, 3290–3295.
[11] Y. Tian, A. Merdasa, M. Peter, M. Abdellah, K. Zheng, C. S. Ponseca, T. Pullerits, A. Yartsev, V. Sundström, I. G. Scheblykin, *Nano Lett.* **2015**, *15*, 1603–1608.
[12] H. Yuan, E. Debroye, G. Caliandro, K. P. F. Janssen, J. van Loon, C. E. A. Kirschhock, J. A. Martens, J. Hofkens, M. B. J. Roeffaers, *ACS Omega* **2016**, *1*, 148.
[13] S. Wang, C. Querner, T. Emmons, M. Drndic, C. H. Crouch, *J. Phys. Chem. B* **2006**, *110*, 23221–23227.
[14] P. Frantsuzov, M. Kuno, B. Jankó, R. A. Marcus, *Nat. Phys.* **2008**, *4*, 519–522.
[15] Y. E. Panfil, M. Oded, U. Banin, *Angew. Chem. Int. Ed.* **2018**, *57*, 4274–4295.
[16] J. Si, S. Volkán-Kacsó, A. Eltom, Y. Morozov, M. P. McDonald, M. Kuno, B. Jankó, *Nano Lett.* **2015**, *15*, 4317–4321.
[17] W. Xu, W. Liu, J. F. Schmidt, W. Zhao, X. Lu, T. Raab, C. Diederichs, W. Gao, D. V. Seletskiy, Q. Xiong, *Nature* **2017**, *541*, 62–67.
[18] D. Shi, V. Adinolfi, R. Comin, M. Yuan, E. Alarousu, A. Buin, Y. Chen, S. Hoogland, A. Rothenberger, K. Katsiev, Y. Losovyj, X. Zhang, P. A. Dowben, O. F. Mohammed, E. H. Sargent, O. M. Bakr, *Science* **2015**, *347*, 519–522.
[19] W. Tian, C. Zhao, J. Leng, R. Cui, S. Jin, *J. Am. Chem. Soc.* **2015**, *137*, 12458–12461.
[20] A. Merdasa, Y. Tian, R. Camacho, A. Dobrovolsky, E. Debroye, E. L. Unger, J. Hofkens, V. Sundström, I. G. Scheblykin, *ACS Nano* **2017**, *11*, 5391–5404.
[21] A. Halder, R. Chulliyil, A. S. Subbiah, T. Khan, S. Chattoraj, A. Chowdhury, S. K. Sarkar, *J. Phys. Chem. Lett.* **2015**, *6*, 3483–3489.
[22] L. Dou, A. B. Wong, Y. Yu, M. Lai, N. Kornienko, S. W. Eaton, A. Fu, C. G. Bischak, J. Ma, T. Ding, N. S. Ginsberg, L. W. Wang, A. P. Alivisatos, P. Yang, *Science* **2015**, *349*, 1518–1521.
[23] W. Tian, R. Cui, J. Leng, J. Liu, Y. Li, C. Zhao, J. Zhang, W. Deng, T. Lian, S. Jin, *Angew. Chem. Int. Ed.* **2016**, *55*, 13067–13071.
[24] B. Wu, H. T. Nguyen, Z. Ku, G. Han, D. Giovanni, N. Mathews, H. J. Fan, T. C. Sum, *Adv. Energy Mater.* **2016**, *6*, 1600551.
[25] T. S. Sherkar, C. Momblona, L. Gil-Escrig, J. Ávila, M. Sessolo, H. J. Bolink, L. J. A. Koster, *ACS Energy Lett.* **2017**, *2*, 1214–1222.
[26] A. Halder, N. Pathoor, A. Chowdhury, S. K. Sarkar, *J. Phys. Chem. C.* **2018**, 122, 15133-15139

## (Graphical) Synopsis

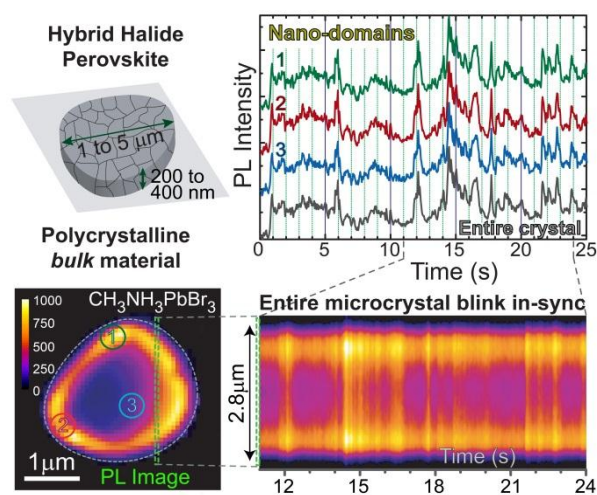

**Blink In-Sync:** A novel phenomenon of spatiotemporally correlated fluorescence intermittency is reported for entire individual organo-metal halide perovskite microcrystals, a bulk polycrystalline material. This observation unambiguously points out to extremely long-range (>μm) communication between photo-generated carriers, and dispels the long standing notion that dimensional nano-confinement is essential to exhibit photoluminescence blinking.

Photoluminescence Blinking beyond Quantum-Confinement: Spatiotemporally Correlated Intermittency over Entire Micron Sized Perovskite Polycrystalline Disks

Nithin Pathoor[1,#], Ansuman Halder[2,#], Amitrajit Mukherjee[1], Jaladhar Mahato[1], Shaibal K. Sarkar[2,*], and Arindam Chowdhury[1,3,*]

[1]Department of Chemistry, and [2]Department of Energy Science and Engineering, Indian Institute of Technology Bombay, Powai, Mumbai 400076, India

*Email: AC: arindam@chem.iitb.ac.in; SKS: shaibal.sarkar@iitb.ac.in

## Supplementary Text

## 1. Experimental Details

### 1.1 Synthesis of methylammonium bromide:

Methylammonium bromide is synthesized by reacting equimolar mixture of methylamine and hydrobromic acid with ethanol as a solvent in a round bottom flask. The solvent is evaporated using a rotary evaporator and the product dissolved in ethanol and recrystallized from diethyl ether. The re-crystallized methylammonium bromide salt is washed repeatedly using diethyl ether and dried in vacuum oven at 60º C for 12 hours.

### 1.2 Preparation of methylammonium lead bromide micro-crystals:

Precursor solution of methylammonium lead bromide ($MAPbBr_3$) is prepared from mixture with 0.22 M methylammonium bromide and 0.2 M lead bromide (sigma-aldrich) in N,N, Dimethylformamide (sigma-aldrich) solvent.[1,2] Here amount of MABr is slightly excess (10%) compared to $PbBr_2$. To form the microcrystals (MCs), the precursor solution is spin coated on a glass cover slip (Fisher Scientific, 25×25×0.17 mm) at 2000 rpm for 60 seconds and annealed at 90°C. Measurements were performed on solid samples formed on the glass coverslips at room temperature (295K) unless otherwise mentioned.

### 1.3. Characterisation of samples:

**1.3.1 UV-Vis Absorption Spectra:** The room temperature solid state absorption spectrum from as-prepared $MAPbBr_3$ MCs was collected in reflection mode with Shimatzu UV-3600 Spectrophotometer using MPC-3100 extension (2D Detector). The absorption spectral profile (**Figure S3**) is in reasonably good agreement with literature, where a clear excitonic peak at 520 nm has been reported.[3]

**1.3.2. Steady State Photoluminescence:** Solid-state fluorescence emission spectra was collected in reflection mode, using a Horiba Fluoromax-4 Spectrofluorometer. The sample, MCs grown on coverslip, was excited at 405 nm at an appropriate angle of excitation so as to avoid reflected scattered light entering the exit slit for detection, in addition to appropriate filters in the detection path. The emission spectrum is quite narrow with a maxima at ~545 nm (**Figure S3**), characteristic of $MAPbBr_3$.[3]

**1.3.3 High Resolution X-ray Diffraction (XRD):** XRD measurements were done with Rigaku Smartlab 9kW HR x-ray Defractometer (θ-2θ scanning mode specific for polycrystalline sample) from 0.5 x 0.5 mm area of thin films grown on glass coverslip. The measurement was performed at 40 kV at 200 mA current. **Figure S1(a)** shows XRD pattern of MAPbBr$_3$ crystalline thin film, in which different peaks are assigned based on literature, and manual calculation confirmed the formation of MAPbBr$_3$ perovskite on the coverslip.

The microcrystals are expected to be polycrystalline since we are using simple solution based sample growth method, on top of an amorphous glass coverslip.[4] We used Debye Scherrer formula to determine dimension of the crystallite related to the broadness of peaks given by, $D = 0.9\lambda/\beta cos\theta$, where λ is the wavelength of x-ray source (0.154 nm), β is the full width at half maxima (FWHM) of a peak (in radians) and θ is the Bragg angle. We determined the average dimension of crystallites corresponding to different prominent peaks and the average size of crystallites is found to be ~30 nm.

**1.3.4 Transmission Electron Microscopy (TEM):** The sample for TEM imaging was prepared by scratching the microcrystal using a razor blade and dispersing it on TEM grid using toluene. TEM measurements were done using Tecnai G2, F30 HR-TEM 300kV instrument. The representative HR-TEM image (**Figure S1(b)**) of MAPbBr$_3$ MC grain show two different orientation of crystal planes implying polycrystalline nature of MCs.

The coexistence of different planes was observed in High Resolution TEM images but not predominantly as in case of other polycrystalline materials. It is noted that in HRTEM images we are limited by dimension of area to be imaged. Hence these microcrystals are indeed polycrystalline in nature and the individual crystallites (ordered domains) are in the order of 30 nanometres dimension.

**1.3.5 Scanning Electron Microscopy (SEM):** SEM imaging were performed using a Zeiss Ultra 55 FE-SEM scanning electron microscope from Oxford Technologies. Quasi circular disc-like structure of microcrystals (**Figure 1(a)**) can be seen in SEM image along with abundant fused MCs, where 2-5 grains are conjoined to each other. Typical size of individual grains varied from ~500 nm to ~10 μm in diameter.

**1.3.6 Atomic Force Microscopy (AFM):** AFM measurements were performed using a Asylum Research MFP 3D BIO instrument in dry condition with as prepared sample (grown on coverslip), in tapping mode. The tip used was AC-240 TS, 70 kHz with force constant 2 N/m. The study of height profile across the MCs indicates considerably flat discs (**Figure S2**) of MAPbBr$_3$ crystals of varying dimensions. Typical cross sectional area of these MCs range from ~0.8-30 μm$^2$ and the thickness (height) vary from 200-400 nm, which is much larger than quantum confined regime, portraying bulk materials.

**1.3.7. Time-resolved Photoluminescence Decay Dynamics:** The ensemble fluorescence lifetime of the as-prepared MAPbBr$_3$ thin film was determined using an IBH Horiba Jobin Yvon FluoroCube time-correlated single photon counting (TCSPC) spectrometer. The solid sample was placed at an angle of 60° with respect to the excitation source of 440 nm from Horiba NanoLED diode laser. The full width at half maximum (FWHM) of the instrument response function (IRF) was ~290 ps. The decay was fitted to tri-exponential function by iterative reconvolution using IBH DAS 6.2 software. We used goodness of fitting parameter as an indicator of best fit for the experimental fluorescence decay. The average lifetime was calculated using the expression: $<\tau> = \frac{A_1\tau_1 + A_2\tau_2 + A_3\tau_3}{A_1 + A_2 + A_3}$, where, $A_1$, $A_2$ and $A_3$ are contributions of lifetime components $\tau_1, \tau_2$ and $\tau_3$, respectively.

**1.4. Epifluorescence video microscopy:** A home built laser epi-fluorescence microscopy setup was used to perform PL video imaging and spatially-resolved emission spectroscopy of single MAPbBr$_3$ MCs under ambient conditions. The details of the experimental setup can be found elsewhere[1,5,6]. For these measurements, a 405 nm laser (LaserGlow, LRD-0405-PFR) was used to excite samples placed on in inverted microscope (Nikon Eclipse 2000U) mechanical stage. A 1.49NA 60x oil immersion objective (Nikon Apo TIRF) was used to illuminate ~1600 μm$^2$ circular area of the sample. The excitation power density could be varied from 0.01-100 W/cm$^2$ using ND filters, however, all data presented in this report were collected at 1W/cm$^2$ power density as measured just below the objective entrance. The emitted light was collected through the same objective lens, passed through a 405 nm dichroic mirror and a 440 nm long-pass filter to remove residual excitation light, and imaged at 35 ms exposure time (5 ms readout time) using an interline CCD camera (DVC 1412AM). All movies were collected as sequence of images in 16 bit format at 25Hz, and analysed after background subtraction with rolling ball radius of 50 pixels using ImageJ 1.50 (NIH).

**1.5. Photoluminescence spectroscopy of single MCs:** Spectrally-resolved PL microscopy was performed using a combination of a narrow slit and a transmission grating (Optometrics, 70 grooves/mm) placed in the emission path before detector.[5,6] For spatially-resolved spectroscopy, samples were excited using a 405 nm laser (LaserGlow, LRD-0405-PFR) with excitation power density of 1W/cm$^2$, and the exposure time was 35 ms (data collected at 25Hz at 16 bit image sequences). The spectral images were processed using imageJ and analysed using OriginLab. The temporal dynamics of single MC emission spectra is shown in the form of a kymogram by plotting the intensity profile along wavelength with illumination time (**Figure S8**), which reveals absence of any spectral diffusion and only abrupt blinking.

## 2. Analyses of photoluminescence video microscopy data:

**2.1. Fluorescence intermittency (blinking) and emission spectra:**

All movies including and spectrally-resolved images were analysed with ImageJ 1.50i (NIH), Origin 8.5 and Matlab R2015b. All intensity-time trajectories of nano-domain or entire MC were extracted using ImageJ, including that for intensity profiles along a line (**Figure 3(d)**, **Figure S11(d)**). Frequency distributions or histograms (**Figures S5, S6, S7, S9, S11**) were constructed using Origin. Spectrally resolved images were analysed by integrating the emission from entire vertical strip along a single MC following previously published procedures[1,5]. Pearson correlation coefficients were evaluated using a program written in Matlab using data sets obtained from ImageJ, and the process was repeated for different nanodomains in the same MC, or for different grains in fused MCs, or various MCs in the ensemble **(Section 2.2, SI)**. Correlation-maps/-matrices were also generated using Matlab.

**2.2. Evaluation of Pearson Correlation Coefficients:**

Pearson cross-correlation provides a statistical expression for extent of correlation between any two data sets ($I^i$ and $I^j$), given by Pearson Correlation Coefficient (PCC),[7]

$$PCC = \frac{N(\sum_t(I^i(t)I^j(t)) - (\sum_t I^i(t))(\sum_t I^j(t))}{\sqrt{\left[N\sum_t(I^i(t))^2 - (\sum_t I^i(t))^2\right]\left[N\sum_t(I^j(t))^2 - (\sum_t I^j(t))^2\right]}}$$

Where, N is the total number of time (*t*) frames (*t* ranges from 1 to N), $I^i$(t) and $I^j$(t) are the intensity values of the *t*$^{th}$ frame of two time-trajectory data sets $I^i$ and $I^j$ respectively.

PCC ranges from 1 to -1, indicating perfect correlation for *1* and anti-correlated for *-1*, while zero indicating two data sets uncorrelated. The data sets with CC value more than 0.5 is considered to be correlated, whereas value more than 0.8 is highly correlated. The data sets with values of <0.4 is considered to have no significant correlation.

It is relevant to note that the value of PCC can be obtained from any two intensity-time trajectories between, (*i*) two regions (grains) within the same fused MC, (*ii*) any one local domain (pixel) and the entire single MC, (*iii*) any two local domains (pixels) within a single MC and (*iv*) two different entire MCs' integrated-intensity traces.

*Details of evaluation of PCC for these categories are provided below:*

(i) Spatially integrated intensity time-trajectories of two entire grains within a fused MCs (**Figure 3a and Figure S11(a)**) was visualized by plotting the two data sets with respect to each other (**Figure 3c and**

**Figure S11(c))** from which the extent of the correlation between intensity fluctuations is quantified. This PCC is assigned as *r* in the manuscript.

(ii) To generate "*Correlation Images*" or "*Correlation Maps*" of individual MCs (**Figure 2(b) and Figure S6(g)**), the spatially-averaged intensity time trajectory of an entire MC ($I^i(t)$) is correlated to intensity time trace of each of the pixel in the MC, one at a time. So, the expression for PCC in these data sets becomes:

$$r_i = \frac{N\sum_t(I_{av}^{MC}(t)I^i(t)) - (\sum_t I_{av}^{MC}(t))(\sum_t I^i(t))}{\sqrt{\left[N\sum_t(I_{av}^{MC}(t))^2 - (\sum_t I_{av}^{MC}(t))^2\right]\left[N\sum_t(I^i(t))^2 - (\sum_t I^i(t))^2\right]}}$$

where, N is the total number of time frames and t ranges from 1 to N, $I_{av}^{MC}(t)$ is the average intensity of the entire crystal at $t^{th}$ frame, $I^i(t)$ is the intensity of the $i^{th}$ pixel at $t^{th}$ frame of one micro-crystal.

The value of the PCC obtained for each pixel ($r_i$) is allotted to the respective pixels, thereby generates a spatial map of PCCs in intensity fluctuations. Spatial distribution of $r_i$ for all pixels within each MC (**Figure S6(h) and S9(e-f)**) are constructed from the correlation images using Matlab. The mean values ($\langle r \rangle$) of $r_i$ (*i.e.*, spatially-averaged $r_i$) obtained from each single-MC distributions are plotted as a function of 125 MCs in **Figure 2(c).** The corresponding distribution of $\langle r \rangle$ for 125 single MCs is shown in **Figure S10(c)**.

(iii) "*Correlation Matrix*" (**Figure 3(e)** and **Figure S11(e)**) is constructed by correlating the intensity time trajectories of each pixel with that of every other pixels of a line strip along a MC. The obtained PCCs of intensity fluctuations between $i^{th}$ and $j^{th}$ pixels' time series are plotted in the form of a 2D (symmetric) matrix (**Figure 3(e)**, **Figure S11(e)**). In this matrix, the diagonal elements represent self-correlation of each pixel (thus, value = 1) and off-diagonal elements represent inter-pixel intensity PCC (defined $r_{ij}$). The color codes in this correlation matrix represent values of $r_{ij}$.

(iv) The extent of external factors such as laser intensity fluctuation may leads to blinking of PL from MCs. To exclude that possibility, which may leads to synchronous fluctuation of different MCs in same movie, 25 spatially isolated MCs from one movie is selected and their spatially averaged intensity-time trajectories are correlated for every pair of MCs. This yields a 2D "*Inter-crystal PCC matrix*" (**Figure S7(b)**), where the diagonal elements represent self-correlation of each MC (thus value = 1) and off-diagonal elements represent inter-MC intensity PCC (assigned the term ***R***). The distribution of ***R*** for 25 isolated MCs (**Figure S7(c)**) constructed from the Inter-crystal PCC matrix shows values ranging from -0.4 to 0.5 with an ensemble average value ($\langle R \rangle$) of ~0.05 and a standard deviation of 0.2. This signifies uncorrelated blinking between any two isolated single MCs of the ensemble. This corroborates that any value of PCC below 0.4 essentially implies spatially uncorrelated blinking behaviors. We noticed however that several pairs of MCs do show value of ***R*** of ~0.4, which we found out is due to slow timescale (0.1-1 Hz) progressive intensity enhancement of these MCs, due to photocuring processes (i.e., continuous enhancement of average base intensity with illumination time) reported previously.[8]

# Supplementary Data (Figures)

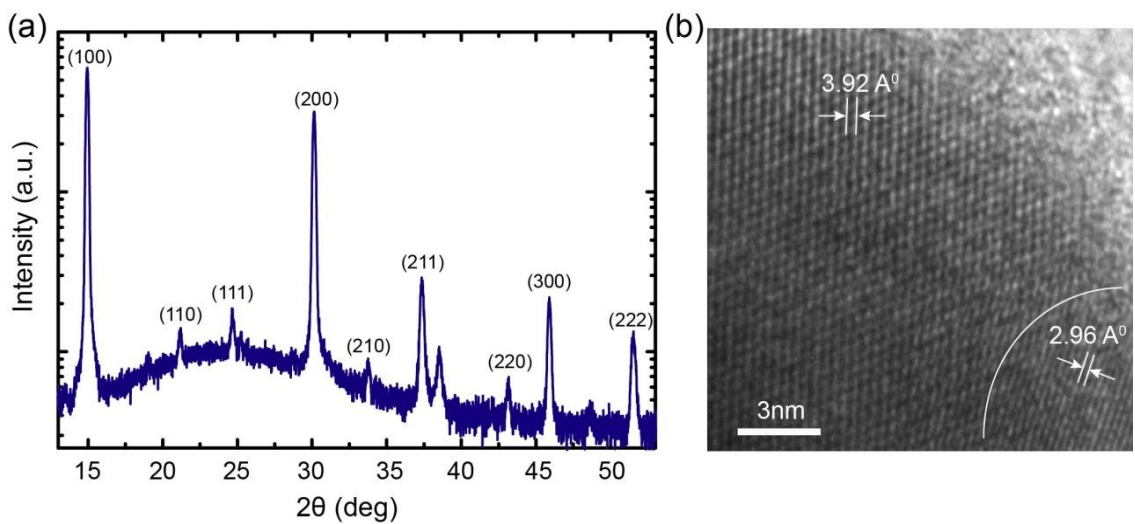

**Figure S1.** (**a**) XRD profile of MAPbBr$_3$ collected from 0.5x0.5 mm$^2$ area of a film grown on coverslip showing the characteristic peaks corresponds to different planes of MAPbBr$_3$, (**b**) HR-TEM image of a small area of MC showing two different planes (*200* and *111* corresponding to 2.96 Å and 3.92 Å inter-planar distances, respectively). Detailed description on the crystallinity of the MCs is provided in the **section 1.3** of SI text.

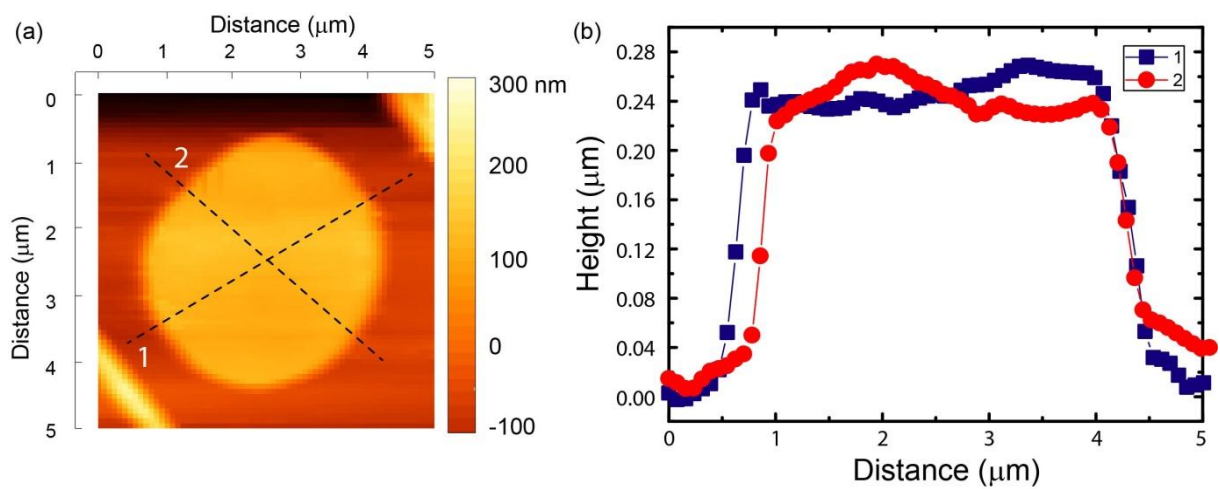

**Figure S2.** (**a**) AFM topographical image of a typical MAPbBr$_3$ microcrystal showing quasi circular flat-disk like shape. (**b**) Height profile along dotted lines in (a) shows thickness of this disk to be about 250 nm. Such a structure with more than 200 nm thickness implies there is no quantum- confinement of carriers in any dimension, and thus represents a "bulk" crystalline material.

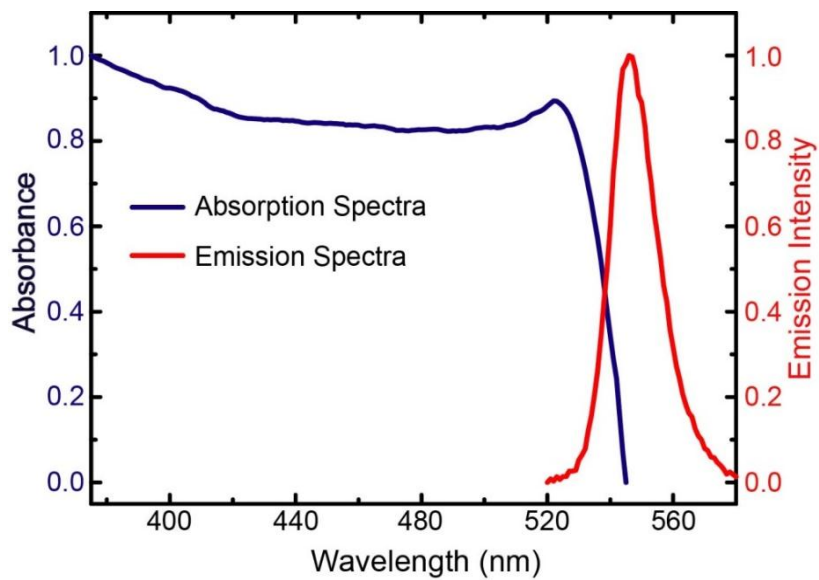

**Figure S3.** Bulk absorption and PL emission spectra of MAPbBr$_3$ MCs in the solid state. The absorption spectrum (blue) shows an excitonic absorption feature at ~520 nm and an intense narrow emission spectrum (red) at 545 nm (consistent with literature[3]). The excitonic binding energy for MAPbBr$_3$ is very close to thermal energy (~25 meV), which implies excitons are short lived and majority of carriers exist as freely diffusing charges (electrons and holes).[9]

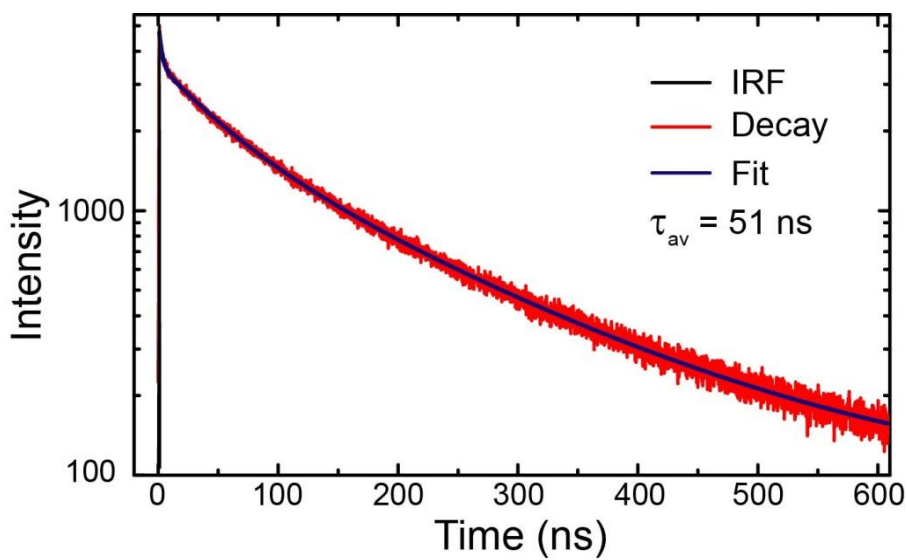

**Figure S4**: Fluorescence decay (red) of MAPbBr$_3$ MCs recorded with $\lambda_{ex}$ = 440 nm and $\lambda_{em}$ = 544 nm, as collected from solid samples using TCSPC. The Instrument Response Function (black) for this setup is 290 ps. We used iterative re-convolution to fit the decay (blue) with tri-exponential function (three components are 2.4ns, 51 ns and 179 ns with contributions of 59%, 17% and 23% respectively); average radiative recombination lifetime is calculated to be ~51 ns.

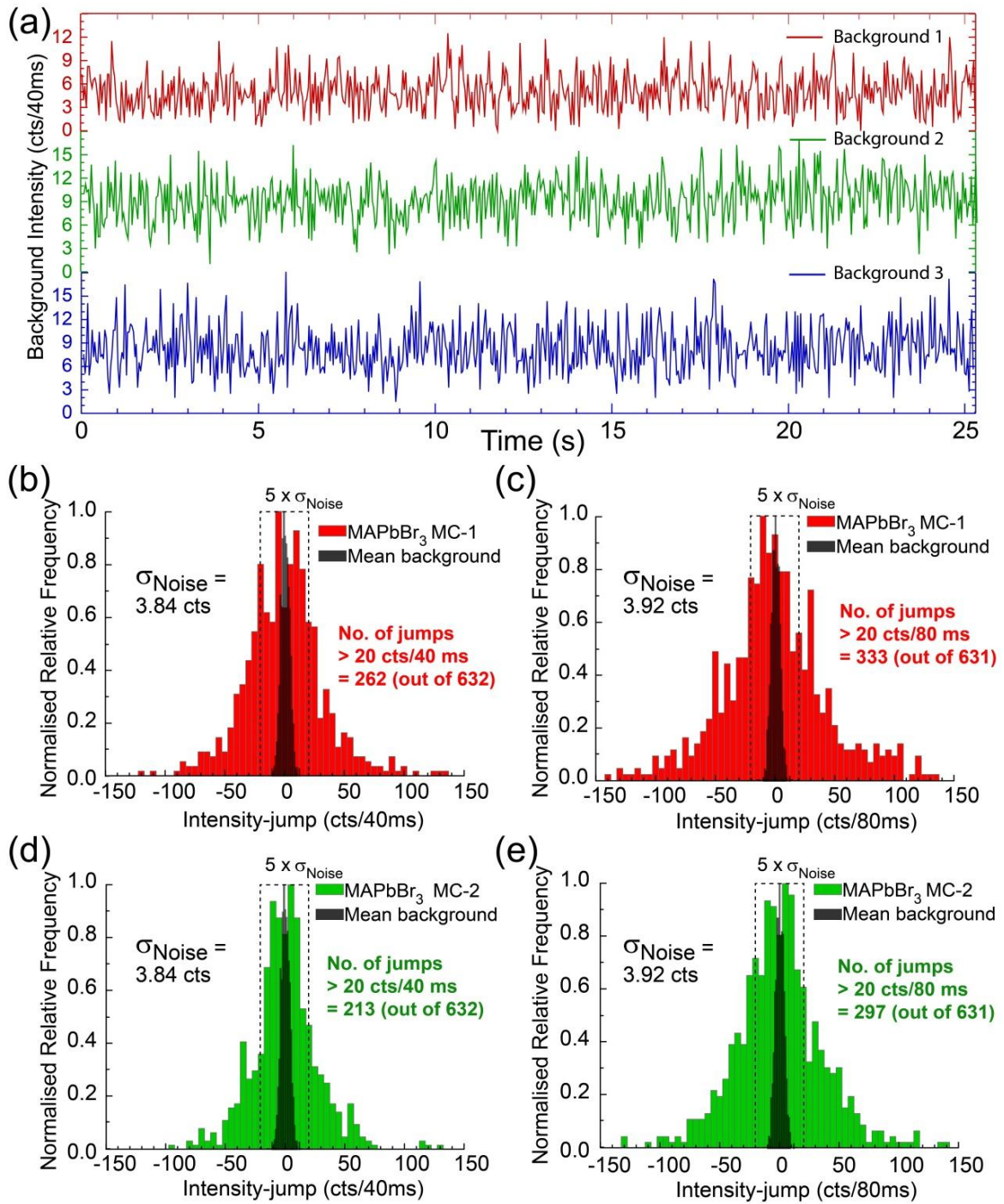

**Figure S5. (a)** Three representative background intensity fluctuations (2x2 pixel area) from different movies, in regions where no MC was present (*i.e.*, under identical excitation power density and exposure times). The abrupt intensity jumps (one and two frames) in each background is analysed separately and distributions constructed from the data of all three traces. **(b-e)** Distribution of intensity jumps of three background traces (grey) and that for MC-1 *(b,c)* and MC-2 *(d,e)* for one-frame *(b,d)* and two-frame *(c,e)* intensity jumps, depicting a large number of blinking events. The standard deviation of intensity fluctuations for three noise traces was found to be ~3.9 counts for both single-frame (40 ms) and double-frame (80 ms) intensity jumps. Intensity jumps only beyond 20 counts per 40 and 80 ms is considered to be blinking events.

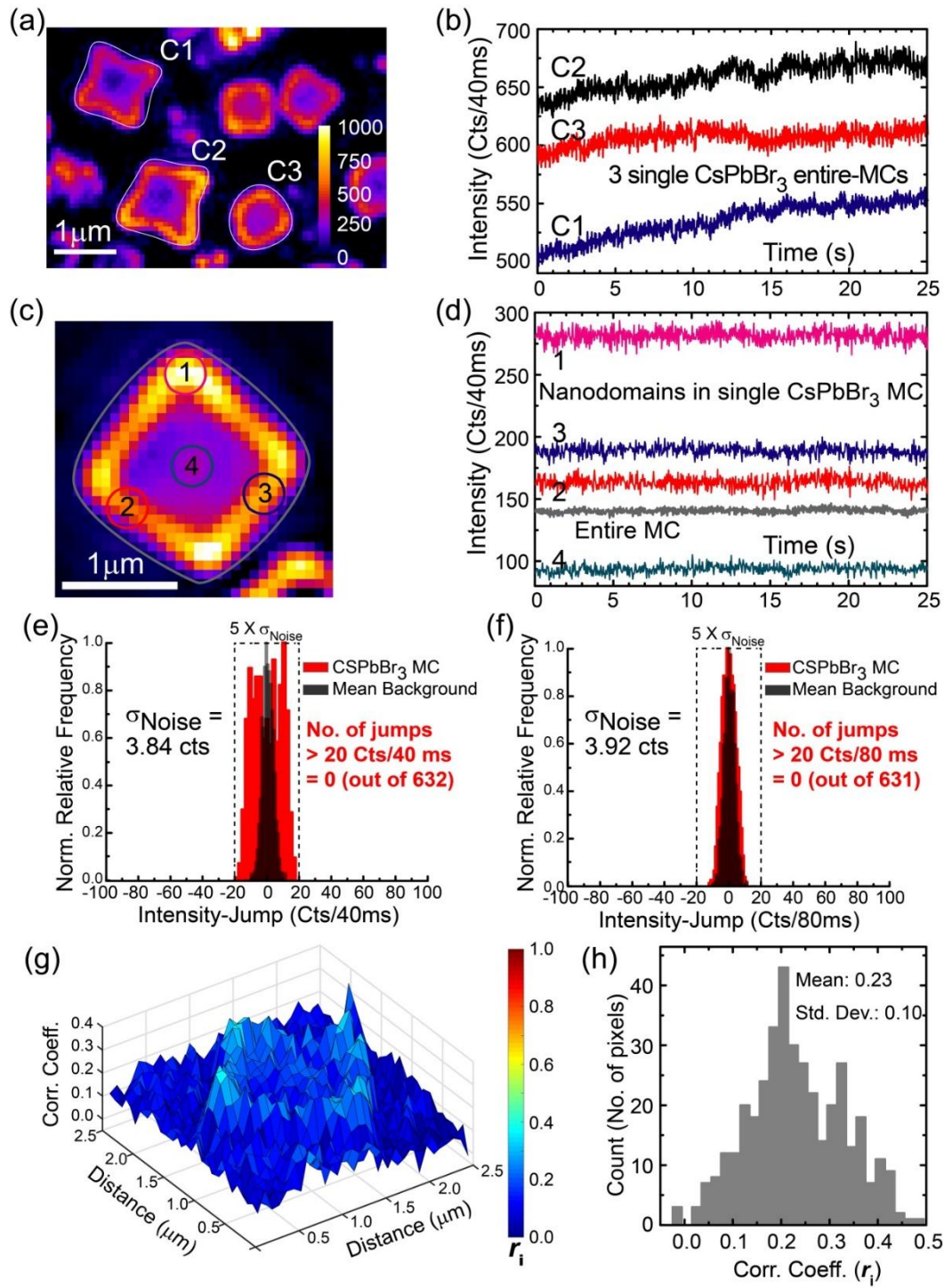

**Figure S6**. **(a)** PL intensity image of all-inorganic $CsPbBr_3$ MCs and **(b)** three representative PL intensity trajectories (under continuous illumination) from entire individual CsPbBr3 MCs, showing lack of any abrupt fluctuations and only slow photobrightening (light soaking effect) **(c)** Blow up PL image of another $CsPbBr_3$ MC and **(d)** intra-crystal PL intensity-time traces of 4 nano-domains of the $CsPbBr_3$ MC (marked in (c)) along with entire MC average intensity trajectory, revealing small amplitude, spatio-temporally uncorrelated fluctuations. **(e-f)** Distribution of intensity jumps for average of three background regions and that for the single entire $CsPbBr_3$ MC for one-frame (e) and two- frame (f) intensity jumps. **(g)** Correlation image (*see section 2.2 (ii), SI*) of the $CsPbBr_3$ MC (shown in c) providing a spatial map of correlation coefficients ($r_i$). **(h)** Distribution of spatial correlation coefficients ($r_i$) within the $CsPbBr_3$ MC (shown in c) revealing nominal spatio-temporal correlation. Relatively high values (0.3-0.45) arise from overall slow modulation of the entire MC. Note that the intensity trajectory shown in *Figure 1d* of manuscript is for another $CsPbBr_3$ MC.

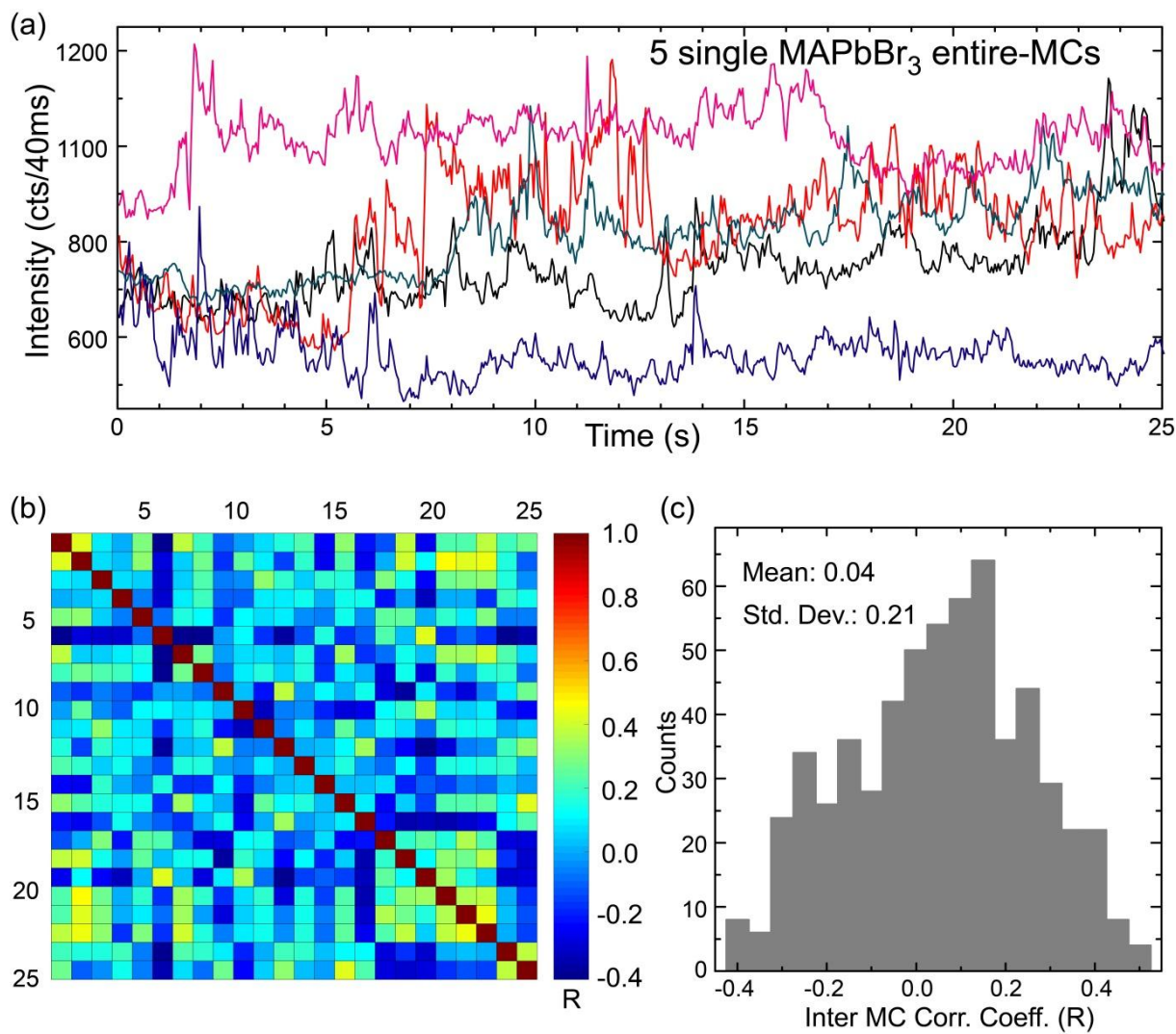

**Figure S7. (a)** Representative emission intensity trajectories (under continuous illumination) for five more segregated MAPbBr$_3$ MCs (chosen from the same movie/data set) showing no two individual crystals have identical blinking behaviors, while the blinking dynamics are quite heterogeneous in the ensemble. **(b)** An inter-crystal PCC (correlation) matrix (***R**, see section 2.2, SI*) constructed from average intensity-time trace of 25 randomly chosen MAPbBr$_3$ MC which exhibit prominent blinking (chosen from same movie/data set), to exemplify lack of significant correlation in inter-crystal blinking dynamics. **(c)** Distribution of inter-crystal intensity cross-correlation coefficients (***R***) (*see Section 2.2, SI*) constructed from ***R*** values in the inter-crystal PCC (correlation) matrix (b). Low ***R*** values (-0.4 to 0.4) with an ensemble average value of 0.04 and standard deviation of 0.2 signify spatially disconnected individual MCs do not undergo temporally correlated blinking.

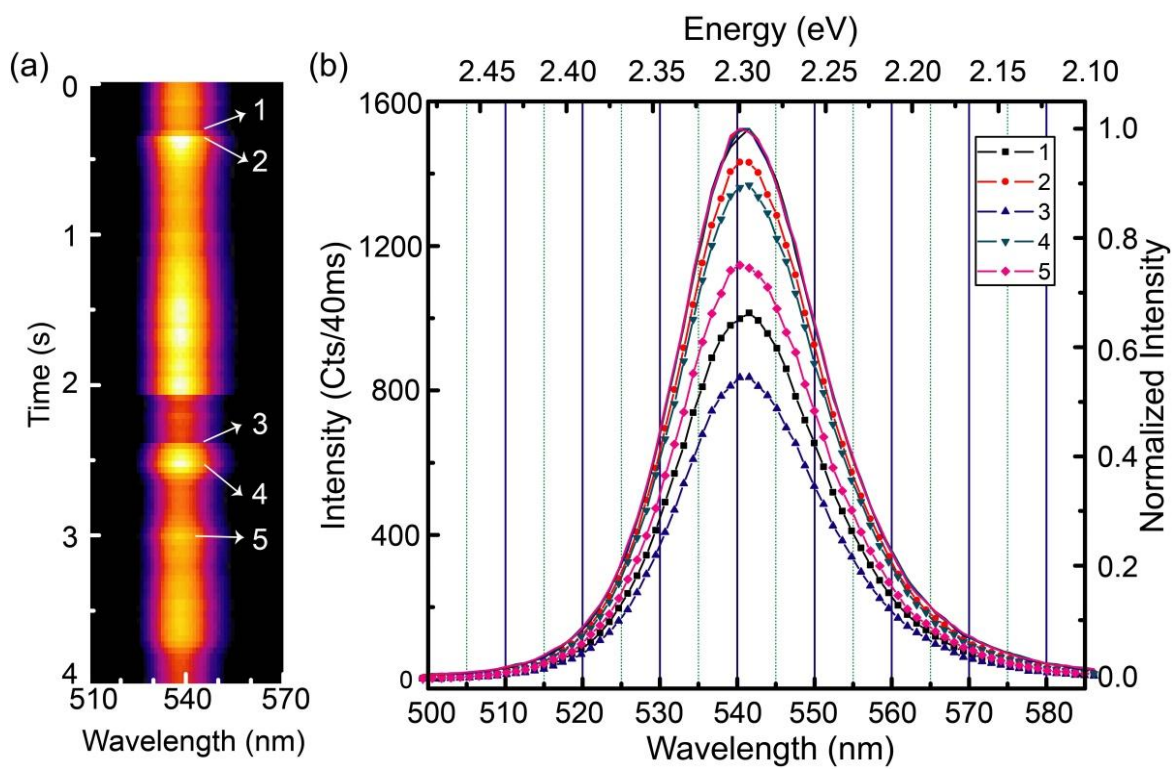

**Figure S8.** Time evolution of emission spectral characteristics for a representative blinking microcrystal of MAPbBr$_3$ **(a)** Temporal evolution of the spatially-average dispersed emission spectra indicating the fluctuation of spectral intensity (and not transition energy) with illumination time. **(b)** A few selected emission spectra of the same microcrystal in various different intensity states indicated by arrows marked 1-5 in (a). The solid curves represents the normalized emission spectra at the five time points, showing near perfect overlapping of spectra, *i.e.*, they have nearly identical peak positions and widths (within our experimental spectral resolution of ~3 nm).

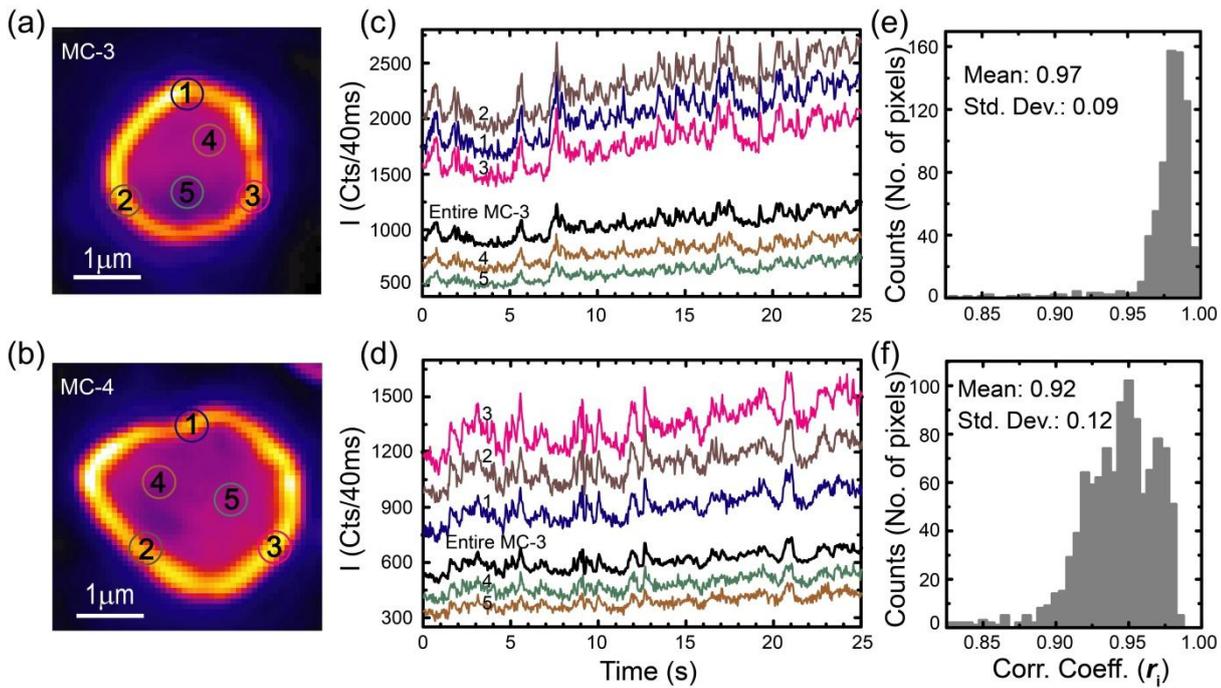

**Figure S9.** (**a-b**) PL intensity images of *MC-3* and *MC-4* (shown in **Figure 2(a)** inset). (**c-d**) Absolute PL intensity (I, in cts/40ms) trajectories of five nano-domains of *MC-3* and *MC-*4, from both edges and interiors, and the average intensity trace for the entire respective MCs. (**e-f**) The distribution of spatial correlation coefficient ($r_i$) (*see section 2.2, SI*) for the pixels where *MC-3* and *MC-4* are present.

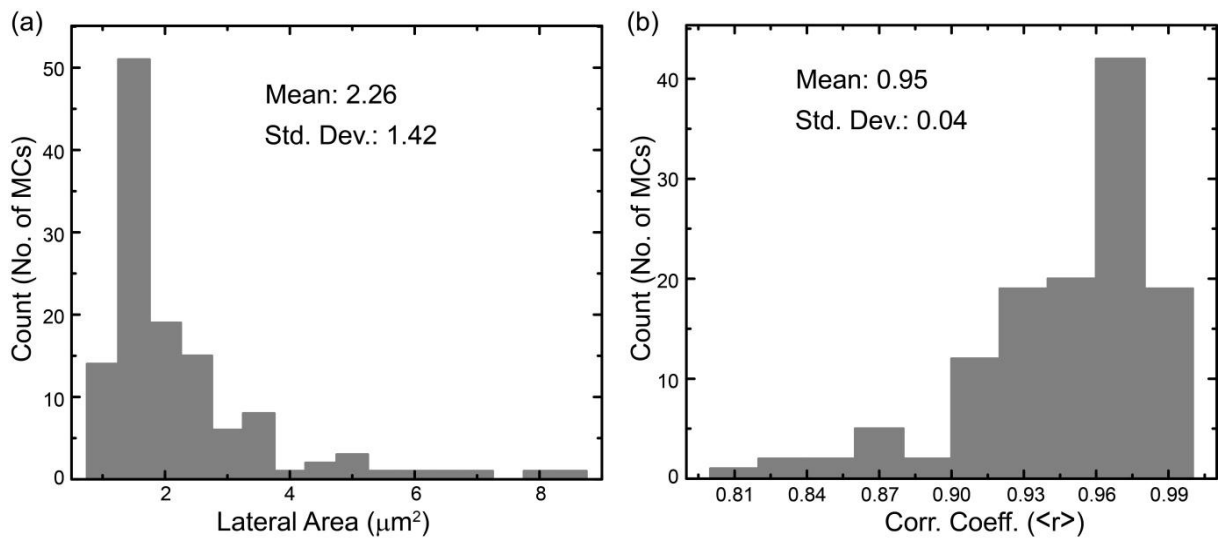

**Figure S10.** (**a**) Distribution of lateral area of 125 microcrystals extracted from PL images by manually calculating the number of pixels over which each MC is located. (**b**) Distribution of spatially averaged intensity correlation coefficients ($\langle r \rangle$) for 125 single MCs whose size distribution provided in (a). Most of the MCs have very high $\langle r \rangle$ above 0.9, and average value is 0.945. (see **Section 2.2** (ii), SI). These distributions were constructed using the data presented in **Figure 2(c)** of the manuscript.

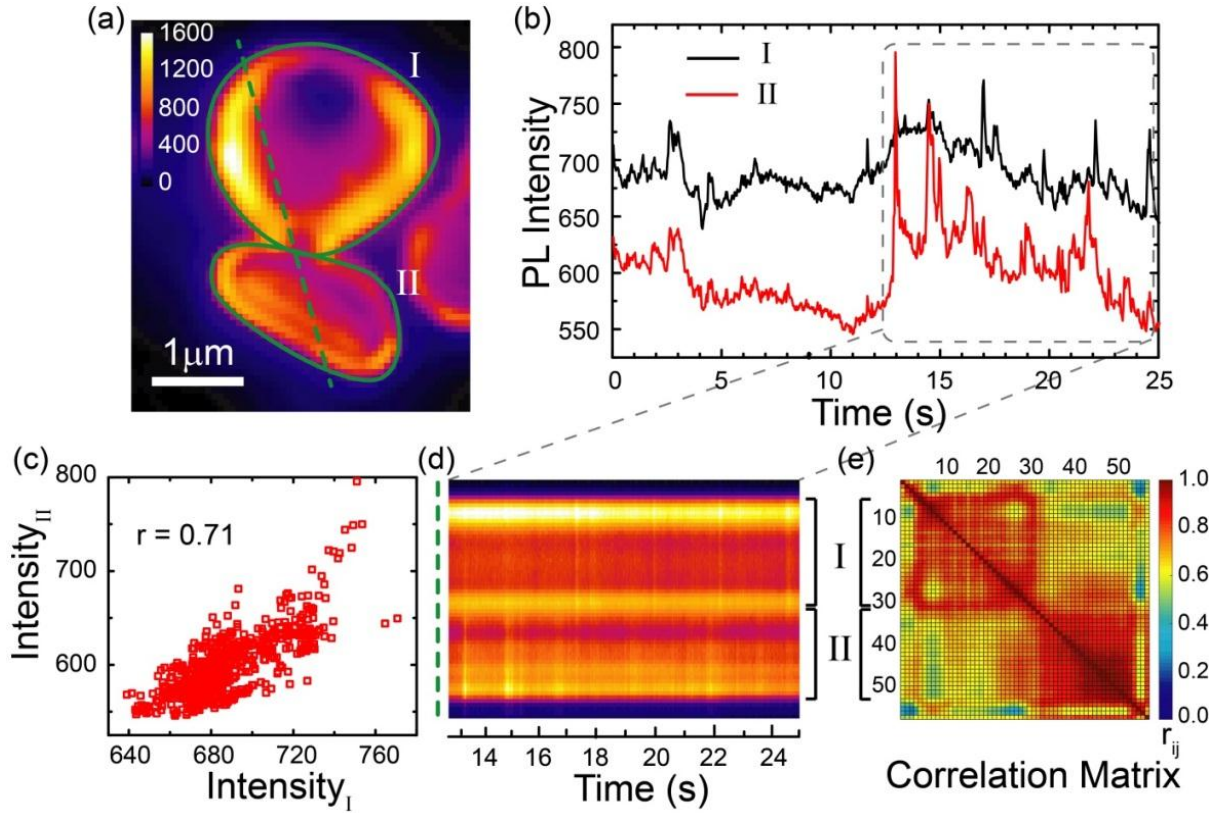

**Figure S11.** Intermittent spatiotemporally-uncorrelated flickering in a fused MC. **(a)** PL image snapshot of fused MC with two grains (I and II), **(b)** Integrated intensity time trace of two individual grains **I** and **II**, where blinking correlation varies over different time windows. (The intensity scale along y-axis is represented in counts per 40 ms). Initially, the two trajectories undergo correlated blinking (up to ~12 s) followed by uncorrelated blinking (12-24 s). However, slow modulation in base intensity is the same for both grains throughout the entire time. **(c)** Integrated intensity of **II** plotted against that of **I** which yields a value of inter-grain correlation coefficient, $r$ = 0.7 (*see section 2.2, SI*) **(d)** Time-evolution of the intensity of a strip along the fused MC and **(e)** the correlation-coefficient ($r_{ij}$) matrix for each pair of pixels (*see section 2.2, SI*) along the line-strip through the fused MC grains. *(c)* and *(d)* demonstrates each of these fused grains undergo concerted blinking on their entirety, while blinking dynamics of individual grains are different from each other (*i.e.*, not very strongly correlated).

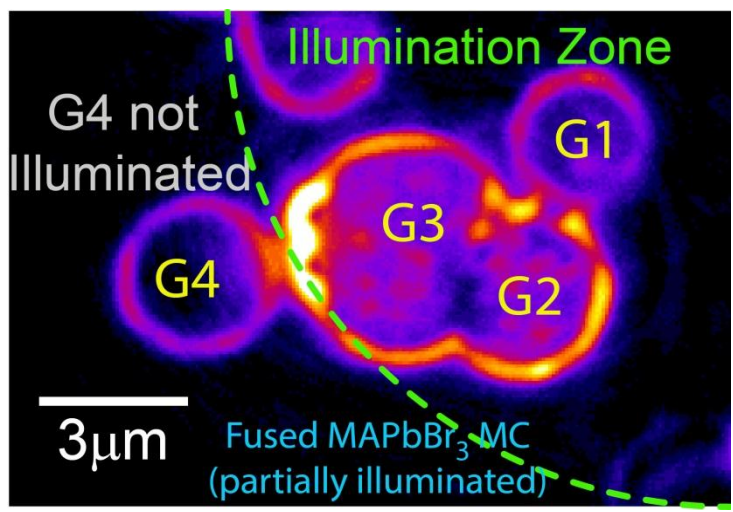

**Figure S12.** Regio-selective illumination of grains GI-GIV in a conjoined MAPbBr$_3$ MC, showing considerable emission emanating from un-illuminated grain (G4) of the fused crystal, indicating the long range migration of charge carriers (>μm) across grain boundary. The measurement was performed under wide-field excitation, the border being marked by a green arc indicating the zone of illumination, and the image was captured using a CCD camera.

# Supplementary References

[1] A. Halder, R. Chulliyil, A. S. Subbiah, T. Khan, S. Chattoraj, A. Chowdhury, S. K. Sarkar, *J. Phys. Chem. Lett.* **2015**, *6*, 3483–3489.
[2] M. Zhang, H. Yu, M. Lyu, Q. Wang, J.-H. Yun, L. Wang, *Chem. Commun.* **2014**, *50*, 11727–11730.
[3] C. Qiu, J. K. Grey, *J. Phys. Chem. Lett.* **2015**, *6*, 4560–4565.
[4] T. S. Sherkar, C. Momblona, L. Gil-Escrig, J. Ávila, M. Sessolo, H. J. Bolink, L. J. A. Koster, *ACS Energy Lett.* **2017**, *2*, 1214–1222.
[5] S. De, A. Layek, A. Raja, A. Kadir, M. R. Gokhale, A. Bhattacharya, S. Dhar, A. Chowdhury, *Adv. Funct. Mater.* **2011**, *21*, 3828–3835.
[6] A. Swarnkar, R. Chulliyil, V. K. Ravi, M. Irfanullah, A. Chowdhury, A. Nag, *Angew. Chemie Int. Ed.* **2015**, *54*, 15424–15428.
[7] J. Lee Rodgers, W. A. Nicewander, *Am. Stat.* **1988**, *42*, 59–66.
[8] Y. Tian, M. Peter, E. Unger, M. Abdellah, K. Zheng, T. Pullerits, A. Yartsev, V. Sundström, I. G. Scheblykin, *Phys. Chem. Chem. Phys.* **2015**, *17*, 24978–24987.
[9] Y. Yang, M. Yang, Z. Li, R. Crisp, K. Zhu, M. C. Beard, *J. Phys. Chem. Lett.* **2015**, *6*, 4688–4692.

# Supplementary Movie (AVI) Captions

All movies presented here are collected with 35 ms exposure time at 25Hz frame rate (including readout time) with a total duration of 25 s, unless otherwise mentioned. All the movies in AVI format are being run at 50 fps. Labels in intensity colour maps represent counts per 40 ms. Note that all these movies were manually enlarged (using ImageJ) by 1.5X compared to the original data, for ease of viewing, *i.e.*, to avoid too much pixilation of the generated AVI files. The length scale bars have been adjusted accordingly.

*Raw data files for movies (16 bit TIFF) are available upon request*

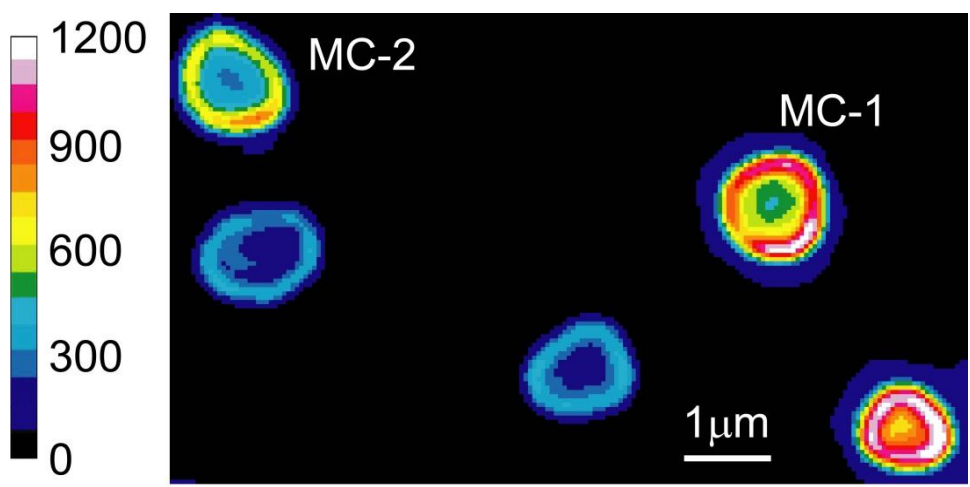

**Movie M1:** PL intermittency of five spatially isolated $MAPbBr_3$ MCs in a 9.3 x 5.5 $\mu m^2$ area (Image sequences shown in **Figure 1b**, for MCs-1 and MC-2, are part of this movie, which has been rotated by 90° for viewing convenience. The colour scale bar represents intensity in counts per 40 ms.

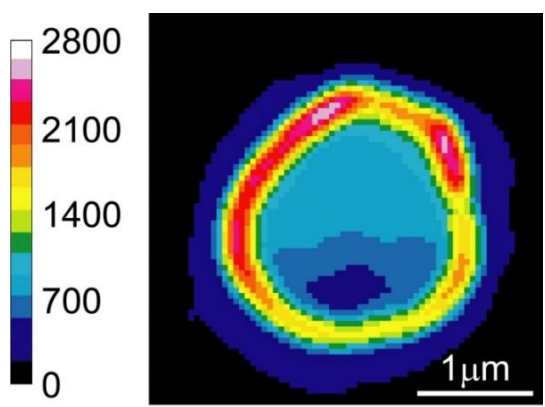

**Movie M2:** PL blinking of *MC-3* (**Figure 3** and **Figure S6**). Video dimensions are 3.4 x 3.4 $\mu m^2$. The colour scale bar represents intensity in counts per 40 ms.

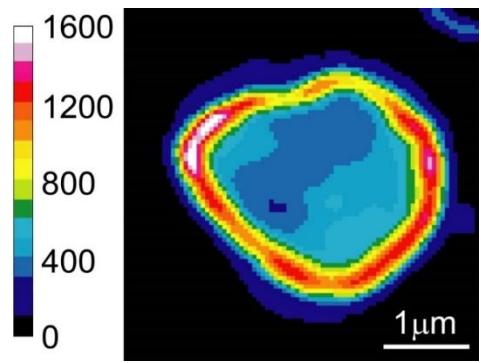

**Movie M3.** PL blinking of *MC-4* (**Figure 3** and **Figure S6**). Dimensions of the video are 4.1 x 4.1 μm$^2$. The colour scale bar represents intensity in counts per 40 ms.

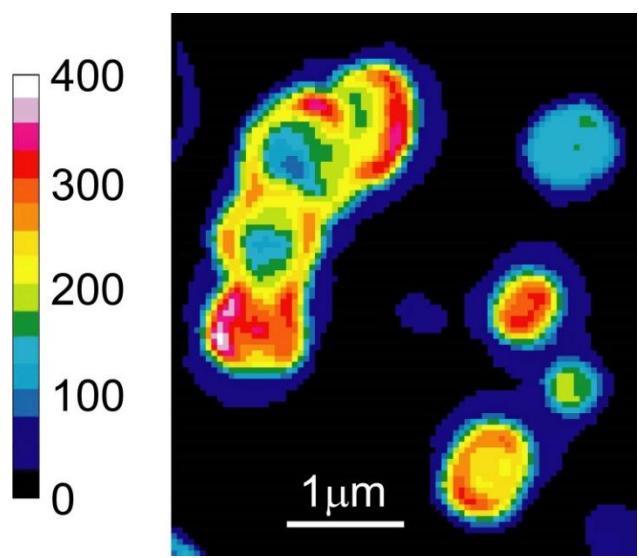

**Movie M4.** PL blinking of an elongated MC with four fused grains **I-IV**, along with few isolated MCs (such as **V**) (see **Figure 3a,** which was cropped from this image). The movie was collected at 32 Hz frame rate and its dimensions are 4.2 x 5.1 μm$^2$. The colour scale bar represents intensity in counts per 33 ms.

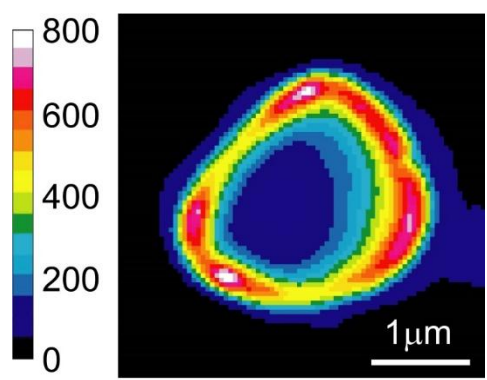

**Movie M5.** PL blinking of the large MAPbBr$_3$ Microcrystal shown in **(Graphical) Synopsis**. Dimensions of the video are 3.8 x 3.8 μm$^2$. The colour scale bar represents intensity in counts per 40 ms.